\documentstyle[12pt]{article} 

\advance\hoffset by -7mm  
\makeatletter
\@addtoreset{equation}{section}
\makeatother

\renewcommand{\title}[1]{\null\vspace{25mm}

\noindent{\Large{\bf #1}}\vspace{10mm}

\noindent {\large By }}
\newcommand{\authors}[1]{\noindent{\large #1}\vspace{3mm}

}
\newcommand{\address}[1]{\noindent #1\vspace{5mm}

}
\renewcommand{\abstract}[1]{\vspace{19mm}

\noindent{\small{\em Abstract.} #1}\vspace{2mm}

} 

\newcommand{\be}{\begin{equation}}
\newcommand{\ee}{\end{equation}}
\newcommand{\g}{{\bf g}}

\newcommand{\M}{{\bf M}}

\newcommand{\A}{{\bf a}}

\newcommand{\mm}{\mu_{0}}

\newcommand{\La}{{\bf L}}
\newcommand{\Ta}{{\bf T}}
\newcommand{\Sa}{{\bf S}}

\begin{document}
\begin{flushright}
hep-th/9609200
\end{flushright}

\title{Sphaleron on $S^{3}$}
\authors{Mikhail S. Volkov}
\address{Institut for Theoretical Physics, 
University of Z\"urich--Irchel,\\ 
Winterthurerstrasse 190, CH--8057 Z\"urich, Switzerland}
\abstract{An exactly solvable sphaleron model in $3+1$
spacetime dimensions is described\footnote{
Contribution to the proceedings of 
Journees Relativistes 96.}}

\section{Introduction}

The notion of {\em sphaleron} refers to the special type
of static classical solution in a gauge field theory
with periodic vacuum structure and broken scale invariance \cite{Manton}.
Specifically, sphaleron relates to the top of the potential barrier
between distinct topological vacua, such that 
its energy determines the barrier height. Sphaleron can
play the important role in the transition processes when
the system interpolates between distinct topological sectors.
Consider a thermal ensemble over one of the
perturbative topological vacua.
Such a system is metastable since the field modes are able to reach the
neighboring topological sectors both via the quantum tunneling
and due to the thermal overbarrier fluctuations.
If temperature is high enough, the latter effect is dominant,
in which case the sphaleron, `sitting' on the top of the barrier, 
controls the transition rate. 
To evaluate the rate of such sphaleron-mediated thermal transitions,
the Langer-Affleck formula is often used \cite{langer}: 
\be                                                    \label{1}               
\Gamma=-\frac{|\omega_{-}|}{\pi}\frac{{\rm Im}Z_{1}}{Z_{0}},
\ee
which relates the probability of the decay of the unstable phase
with the imaginary part of the free energy.
Here $Z_{0}$ and $Z_{1}$ are the 
partition functions for the small fluctuations around the vacuum 
and the sphaleron, respectively. Since the sphaleron has one unstable mode
whose eigenvalue $\omega_{-}^{2}<0$, the quantity $Z_{1}$ is
purely imaginary. To compute $\Gamma$ at the one-loop level
is usually rather difficult. The problem becomes especially hard
in the standard model case, where the  sphaleron solution itself is
known only numerically. That is why the other sphaleron models for which
$\Gamma$ can be evaluated exactly have been investigated
\cite{shap}, \cite{mottola}, however these models 
exist only in $1+1$ spacetime dimensions.

\section{The sphaleron on $S^{3}$}

To find an exactly solvable sphaleron model in $3+1$
dimensions \cite{main}, we consider the theory of a {\em pure}
non-Abelian $SU(2)$ gauge field in
the static Einstein universe $(M,\g)$,
where $M=R^{1}\times S^{3}$, and the metric is 
\be                                                          \label{2}
ds^{2}=\A^{2}\{-d\eta^{2}+ d\xi^{2}+sin^{2}\xi
(d\vartheta^{2}+sin^{2}\vartheta d\varphi^{2})\};         
\ee
here $\A$ is a constant scale factor. Consider the following
$SU(2)$-valued function on the $S^{3}$:
\be                                                          \label{3}
U=U(\xi,\vartheta,\varphi)=
\exp\left\{- i \xi\ n^{a}\tau^{a}\right\},                     
\ee
where $n^{a}=(\sin\vartheta\cos\varphi, \sin\vartheta\sin\varphi,
\cos\vartheta)$ and $\tau^{a}$ are the Pauli matrices.
This function defines the mapping
$S^{3}\rightarrow SU(2)$ with the unit winding number. Using
$U$, we construct the following sequence of the static gauge field
potentials:
\be                                                          \label{4}
A[h]=i\frac{1+h}{2}UdU^{-1},               
\ee
where the parameter $h\in [-1,1]$. When $h=-1$ this field vanishes,
whereas for $h=1$ it is a pure gauge whose winding number is one,
by construction. Thus fields (\ref{4}) interpolate between
the two distinct topological vacua, and the energy
\be                                                          \label{5}
E[h]=\int T^{0}_{0}\sqrt{^{3}\g}d^{3}x=
\frac{3\pi^{2}}{g^{2}\A}(h^{2}-1)^{2}            
\ee
has the typical barrier shape --- it vanishes at the vacuum
values of $h$, $h=\pm 1$, and reaches its maximum in between,
at $h=0$; ($g$ in (\ref{5}) stands for the gauge coupling constant).
The top of the barrier relates to the field configuration
\be                                                      \label{6}
A^{(sp)}=
\frac{i}{2}\ UdU^{-1},
\ee
which obeys the Yang-Mills equations and therefore
can be naturally called sphaleron. It is worth noting that the sphaleron
configuration consists of the gauge field alone. The violation
of the scale invariance in this case is provided by the background
curvature. Since the spacetime geometry is $SO(4)$-symmetric, 
the sphaleron inherits the same symmetries, such that, for instance, the
energy-momentum tensor for the field (\ref{6}) has the manifest
$SO(4)$-symmetric structure.

\section{The sphaleron transition rate}

Our main task is to compute the transition rate (\ref{1}) for the
sphaleron solution (\ref{6}). We pass to the imaginary time $\tau$
in the metric (\ref{2}) and impose the periodicity condition,
$\tau\in [0,\beta]$. Let us introduce $A^{\{j\}}_{\mu}=j A^{(sp)}_{\mu}$,
which corresponds to the vacuum of the gauge field for $j=0$ and
to the sphaleron field for $j=1$. Next we consider small fluctuations
around the background gauge field:
$A^{\{j\}}_{\mu}\rightarrow A^{\{j\}}_{\mu}+\phi_{\mu}$.
Notice that we assume the spacetime metric (\ref{2}) to be fixed
and therefore do not take into account the
gravitational degrees of freedom.
The partition functions $Z_{j}$ are then given
by the Euclidean path integral over $\phi_{\mu}$. To compute the
integral, we impose the background gauge condition and use the
Faddeev-Popov procedure. The result is \cite{main}:
\be                                                \label{7}
Z_{j}=\exp (-S[A^{\{j\}}])\ {\cal N}
\frac{Det'(\hat{M}^{FP}_{j}/\mm^{2})}
{\sqrt{Det'(\hat{M}_{j}/\mm^{2}})},
\ee
where $S$ is the Euclidean action, the factor ${\cal N}$ is due to
the zero and negative modes whereas $Det'$ has all such modes omitted,
$\mm$ is an arbitrary normalization
scale, and the fluctuation operators are
\be                                                  \label{8}
\hat{\M_{j}}\phi^{\nu}=-D_{\sigma}D^{\sigma}\phi^{\nu}
+R^{\nu}_{\sigma}\phi^{\sigma}+
2i[F^{\nu}_{\ \sigma},\phi^{\sigma}],\ \ \ \ \
\hat{\M_{j}}^{FP}\alpha=-D_{\sigma}D^{\sigma}\alpha.
\ee
Here $D_{\mu}=\nabla_{\mu}-i[A^{\{j\}}_{\mu},\ \ ]$ is the covariant
derivative,
$R^{\nu}_{\sigma}$ is the Ricci tensor for the geometry (\ref{2}),
$F^{\nu}_{\ \sigma}$ is the gauge field tensor for $A^{\{j\}}_{\mu}$,
and $\alpha$ is a Lie algebra valued scalar field.

To find spectra of these operators, we introduce
the 1-form basis $\{\omega^{0},\omega^{a}\}$ on the spacetime manifold,
where $\omega^{0}=d\tau$, and $\omega^{a}$ are the
left invariant 1-forms on $S^{3}$. It is convenient to expand
the fluctuations as
$\phi=(\phi^{0}_{p}\omega^{0}+\phi^{a}_{p}\omega^{a})\tau^{p}/2$.
Let $e_{a}$ be the
left-invariant vector fields dual to $\omega^{a}$, such that
$\La_{a}=\frac{i}{2}e_{a}$ are the $SO(4)$ angular momentum
operators. We introduce also spin and isospin operators as follows:
$\Sa_{a}\phi^{b}_{p}=\frac{1}{i}\varepsilon_{abc}\phi^{c}_{p}$ and
$\Ta_{p}\phi^{a}_{r}=\frac{1}{i}\varepsilon_{prs}\phi^{a}_{s}$.
As a result, the fluctuation operators (\ref{8}) can be expressed
entirely in terms of the operators $\La_{a}$, $\Sa_{a}$ and $\Ta_{p}$,
such that the spectra can be explicitly obtained by the purely
algebraic methods \cite{main}. 
All of the eigenvalues are
positive except for the following ones:
the sphaleron fluctuation operator $\hat{\M_{1}}$ has one
negative mode, whereas the vacuum operators $\hat{\M_{0}}$ and
$\hat{\M^{FP}_{0}}$ have three zero modes each. It is worth noting
that, since the sphaleron field configuration is $SO(4)$ invariant,
the sphaleron itself does not have zero modes at all (in the
background gauge imposed).

The next step is to compute the products of the eigenvalues
to evaluate the determinants in (\ref{7}). For this, zeta
function regularization scheme has been used. Omitting all
technical details given in \cite{main}, the resulting
expression for the transition rate can be represented in the
following form:
\be                                                   \label{9}
\Gamma=
\frac{1}{8\sqrt{2}\pi^{2}\sin(\beta/\sqrt{2})}
\exp\left\{-\frac{3\pi^{2}}{g^{2}(\A)}\beta-{\cal E}_{0}\beta
-\beta(F_{1}-F_{0}) \right\}.                                                 
\ee
In this expression, the prefactor in the right hand side is the
overall contribution of zero and negative modes. 
$3\pi^{2}\beta/g^{2}(\A)$ is the
Euclidean action of the sphaleron, where the gauge coupling
constant receives the quantum correction due to the
scaling behavior of the functional determinants:
\be                                                \label{10}
\frac{1}{g^{2}(\A)}=\frac{1}{g^{2}(\A_{0})}
-\frac{11}{12\pi^{2}}\ln\left(\frac{\A}{\A_{0}}\right).
\ee
Here we have replaced $g$ by $g(\A_{0})$, where $\A_{0}=1/\mm$.
This expression agrees with the renormalization group flow,
such that it does not depend on the scale $\A_{0}$ if $g(\A_{0})$ is
chosen to obey the Gell-Mann-Low equation. To fix the scale, we assume that
the value of $g(\A_{0})$ is determined by the physical
temperature, $T(\A_{0})=1/\beta\A_{0}$, and use the QCD data:
\be                                                 \label{11}                
T(\A_{0})=100\ {\rm GeV},
\ \ \ \ \ \ \frac{g^{2}(\A_{0})}{4\pi}=0.12.
\ee
One can assume that the weak coupling region extends up to some
$\A_{max}\sim 10\div 100\A_{0}$.
The next term in (\ref{9}), ${\cal E}_{0}$, is  the
contribution of the zero field
oscillations, that is, the Casimir energy. This quantity can be
computed exactly \cite{main}, the numerical value is
${\cal E}_{0}=-1.084$.
The contribution of the thermal degrees of freedom in
(\ref{9}) is
$$
\beta (F_{1}-F_{0})=
4\ln(1-e^{-\beta})
+2\sum_{\sigma=0,1,2}\ \sum_{n=3}^{\infty}(n^{2}-\sigma^{2})
\ln(1-e^{-\beta\sqrt{n^{2}+\sigma^{2}-3}})-
$$
\be                                                      \label{12}
-6\sum_{n=2}^{\infty}(n^{2}-1)
\ln(1-e^{-\beta n}).                                
\ee
Altogether Eqs.(\ref{9})-(\ref{12}) provide the
desired solution
of the one-loop sphaleron transition problem. The numerical
curves of $\Gamma(\beta)$ evaluated according to these formulas
for several values of $\A$ are presented in \cite{main}.
This solution makes sense under the following assumptions:
\be                                                   \label{13}
\A\leq\A_{max},\ \ \ \ \ \
\frac{1}{\sqrt{2}\pi}<\frac{1}{\beta}\ll\frac{3\pi^{2}}{g^{2}(\A)}.
\ee
The first condition is the the weak coupling requirement.  
When the scale factor $\A$ is too large, the running coupling 
constant (\ref{10}) becomes big (confinement phase), and 
the effects of the strong coupling can completely change the 
semiclassical picture. That is why our solution can be trusted only
for the small values of the size of $\A$. The other condition 
in (\ref{13}) requires that the thermal fluctuations are 
small compared to the classical sphaleron energy,  such that 
the perturbation theory is valid.


\begin{thebibliography}{99}
\bibitem{Manton} N.S.Manton, {\it Phys.Rev.} {\bf D 28} (1983) 2019.

\bibitem{langer} L.S. Langer, {\it Ann.Phys.} {\bf 41} (1967) 108;
I. Affleck, {\it Phys.Rev.Lett.} {\bf 46} (1981) 388.

\bibitem{shap} A.I. Bochkarev, M.E. Shaposhnikov,
{\it Mod.Phys.Lett.} {\bf A 2} (1987) 417.

\bibitem{mottola} E. Mottola, A. Wipf, {\it Phys.Rev.}
{\bf D 39} (1989) 588.

\bibitem{main} M.S. Volkov, {\it Computation of the winding number
diffusion rate due to the  cosmological sphaleron}, hep-th/9604054,
to appear in {\it Phys.Rev.} {\bf D}.


\end{thebibliography}
\end{document}